# Making Internal Software Startups Work: How to Innovate Like a Venture Builder?

Anastasiia Tkalich[1[0000-0001-7391-4194]], Nils Brede Moe[1[0000-0003-2669-0778]]

and Rasmus Ulfsnes[1[0000-0002-4966-8242]]

[1]SINTEF Digital, 7034 Trondheim, Norway
{anastasiia.tkalich, nils.b.moe, rasmus.ulfsnes}@sintef.no

**Abstract.** With the increasing availability of software usage and the influence of the Lean Startup mindset, more and more companies choose to innovate through internal software startups. Such startups aim at developing new business models while at the same time relying on the resources from the companies where they emerged. The evidence from both researchers and practitioners indicates that driving internal software startups is challenging. This paper seeks to address this problem by asking the research question: *how to make internal software startups work?* We examined a unique case of a venture builder – a company primarily focusing on building internal software startups and launching them as independent companies. Applying a Grounded Theory approach, we analyzed data on four internal software startups at the case company. The results suggest that four strategies drive the examined startups: *cultural*, *financial*, *personnel*, and *venture arrangement*. We interpret our results by drawing on earlier literature on intrapreneurship and internal ventures and suggest four recommendations to succeed with internal software startups: 1) establish shared arenas for the employees; 2) provide necessary resources for experimentation in the initial phase and increase them incrementally; 3) build up in-house product management competence through coaching, and 4) harness employees´ own motivation to develop their own ideas.

## 1 Introduction

Innovation is seen as the most critical company capability for company performance[1], which is clearly seen in startups that even with limited resources and capabilities are often able to outperform established organizations. Anthony et al. [2] demonstrated that the average lifespan of a large company is continuously decreasing (based on the S&P500 index of the 500 largest companies). The short lifespan makes innovation not only essential but vital for many companies. At the same time, innovation through traditional R&D units that are primarily used to improve and modify the existing products is not always a solution for innovation because R&Ds have not been sufficiently productive in the last two decades [3]. Creating space, allocating resources, and systematically fostering an innovation culture have thus been solutions for many companies. Giving a set percentage of time for employees to work on projects of their choosing has been implemented by companies such as Google, 3M, Atlassian [4, 5]. Another strategy is to allocate dedicated days (hackathons) where developers work together to develop new products and services [4, 6]. Lastly, an *internal startup* is yet another approach that utilizes the startup concept to foster innovation. For example, Lean startup is a popular way for established companies to create innovative software products [7, 8]. Given that software, data, and AI are becoming increasingly common instruments to innovate, it makes sense to differentiate *internal software startup* as a concept on its own, as we do in this study.

Implementing the Lean startup approach in an established company is more challenging than in a stand-alone startup. This is because internal startups, including *internal software startups*, are bound to the current business strategy of the parent company, which is often supported by rigid bureaucratic structures [9]. Bureaucracy and authoritarian aspects of the parent company constrain the internal startups and deprive them of autonomy, which is their fundamental need. When the startup depends too much on the parent company regarding decisions and resources, the startup's speed and flexibility are reduced [10, 11]. Therefore, many internal startups fail or transition outside the parent companies, as in Lokki, an internal software startup in F-secure [12].

Nevertheless, internal software startups remain a tempting approach for numerous companies who seek practical advice on succeeding. The goal of this study is thus twofold: 1) to provide insight into how to best operate internal software startups and 2) to acquire knowledge and vocabulary to better understand internal software startups. Therefore, we in this study ask the research question:

*What makes internal software startups work?*

To answer this research question, we analyze practical experience from a company specializing in creating and developing internal software startups. This is an extension of the earlier preliminary results [10] that builds on additional data and more rigorous data analysis. We believe this unique case can provide valuable insight for companies that need to grow, nurture, and coordinate several internal software startups; and researchers studying internal startups. The paper is structured in the following way. The next chapter gives an overview of the literature that sheds light on internal software startups. Chapter 3 outlines our research approach and the case context. We present the findings of the study in Chapter 4 and discuss the research question in Chapter 5.

## 2 Related work

*Internal software startup* as we understand it in this paper is built upon the concepts of a standalone *software startup* and extrapolated to the context of entrepreneurship within an established company (*internal startups*). We will now elaborate on this view by drawing on earlier literature. *Startups* are generally seen as temporary organizations with little or no operative history that intend to develop scalable and repeatable business models [13]. According to Eric Reis, a startup is a human institution designed to create new products and services under extreme uncertainty [14]. Since software with an increased focus on data and AI are key innovative differentiators [15], it is important to distinguish *software startup* as a concept on its own. *Software startups* are risk-taking and proactive initiatives that develop *software products* under highly uncertain conditions by constantly searching for repeatable and scalable business models [16, 17]. Such startups utilize software-enabled technologies to rapidly grow and disrupt markets relying on continuous and agile processes. While having the edge over established companies in agility and speed, software startups also suffer from challenges in their own right. For example, in the early stages, the startups may struggle to develop the features that interest customers, build the entrepreneurial team, and find financial resources [18].

Although the notion of an *internal startup* is relatively new in software engineering, the issue of entrepreneurship within existing companies has been scrutinized by management literature for decades. In her comprehensive literature review Lengnick-Hall [19] outlined four different routes to corporate entrepreneurship:

- *Formal research and development* – units consisting of specialists who focus on innovation and the creation of knowledge as their primary objective;
- *Intrapreneurship and internal venture* – individual employees/teams of employees working beyond their normal responsibilities to develop a specific product or process;
- *External joint ventures* – two or more firms pooling their resources to achieve innovation;
- *Acquisition* – innovation through the purchase or stock merger of existing firms.

There are several important distinctions between R&D and internal venture, according to Bart [20]. First, an internal venture is dedicated to the outcome of the venture (e.g. a product) and not to innovation in general. Further, participants in the internal venture are responsible for all phases of the innovation process while at the same time continuing to fulfill their other responsibilities in the company. Gradually, such employees can be fully reassigned to the ventures.

By combining the concepts of *software startups* and *internal ventures/intrapreneurship*, one can define *internal software startups.* These are initiatives that are developed inside of parent companies to achieve software product innovation [11]. Within software development, it is crucial to reduce the uncertainty related to developing a new software product, which gave rise to the concept of Lean startup. Lean startup [21] was suggested as an approach to iteratively develop both the customer problem and its solution through the build-measure-learn loop. A startup should build a product, measure how the customers respond, and learn whether to pivot or continue, which all together allows to better formulate and solve the customer problem. Internal startups do not have to remain in the company, for example, a *spin-off* is an internal startup developed by an employee and with technology from the parent company and later launched as an independent firm [22].

Building upon the summarized above, we suggest the following definition of an *internal software startup*:

- a temporary organization with short or no operative history
- that searches for scalable and repeatable business model
- and develops new *software* products or services under extreme uncertainty
- while relying on technology developed by individual employees/teams of employees working beyond their normal responsibilities in the parent firm

Some companies specialize in building up internal software startups and are thus known as *venture builders* (aka venture factories, company builders, or venture studios) [23, 24]. Venture builders provide support to their startups, such as identifying business ideas, building teams, finding capital, and governing [24]. In return, they receive equity and certain control over the stakes in the emerging startups [23]. Unlike incubators and accelerators, venture builders operate as permanent organizations and are deeply involved with their startups up until they exit [24]. These traits make venture builders a valuable source of knowledge on how to best initiate and operate internal software startups. However, research on *intrapreneurship/internal ventures* in the context of *internal software startups* driven by *venture-building* companies seems scarce. Our study seeks to address this gap of knowledge.

## 3 Research approach

To answer the research question, we collected data from four internal software startups at Iterate, which can be described as a *venture builder* [23]. This section describes our case context and our research approach.

### 3.1 Case context

Iterate is a technology and investment company in digital product development. The 80 employees are software engineers, designers, product managers business developers. Iterate's approach to development and innovation is continuous experimentation. For the past ten years, the company has been named Norway's top three best workplaces (Great Place to Work), and in 2020 Iterate was listed among the 100 best workplaces for innovators in the world [25].

Iterate has three business areas: 1). **Investments**: incubation of employees' own ideas or helping external startups build their product, where Iterate enters as an investor and technologist/design partner. 2). **Consulting**: System development and design on behalf of others (combinations of Java, JavaScript, Clojure, and Scala with DevOps and Continuous Deployment), including corporate ventures. 3). **Software as a Service**: Software tools for innovators, built on insights gained in the other business areas. Iterate builds what they call an ecosystem for innovation, where employees can alternate between working in client assignments and developing their own ideas. In such a way the company acts as an investment fund, technology partner, and employer. Many of the internal startups (ventures) have an environmental profile and cover many domains, such as artificial intelligence for maritime surveillance (Vake), locally produced clothing (Woolit), sustainable food production (Dagens,) and rehabilitation in healthcare (Flow Technologies). Table 1 shows an overview of the startups that we examined. All of these startups completely relied on software for their value creation, which allows them to categorize them as *software ubiquity* [26].

**Table 1.** Profiles of the internal startups

| | Startup 1 | Startup 2 | Startup 3 | Startup 4 |
|---|---|---|---|---|
| Current/maximum team size | 5 (executive team) | 5 | 4 | 5 |
| Composition | CTO, CEO, CFO, developer, DPO, excl. a sales team | Developers and designers | Founder, developers, designer | CEOs and developers |
| Product type | Application for rehabilitation of people with chronic diseases | Online calculation tool | Application for household management | Artificial intelligence for maritime surveillance |
| Timeframe | 2017 - nowadays | 2020 - nowadays | 2020-2021 | 2018 - nowadays |

### 3.2 Data collection and data analysis

Iterate was familiar to us, as we had studied them for five years concerning other innovation topics and found them interesting and suitable for this study. Iterate was part of a research project on software development and innovation in turbulent contexts. The data collection on the internal software startups (the embedded units of analysis) was performed between October 2020 and August 2021. We conducted 7 interviews with a total of 9 participants (see Table 2), and collected documents and notes from several meetings with the company´s representatives (e.g. kick-off meetings on the research project, feedback sessions on the emerging results). For the semi-structured interviews, we used interview guides that slightly differed for the startup founders and the executives. We asked the startup founders questions like *Please, tell us the story of* *your startup*, *what was important for succeeding? Which support do you receive from Iterate? What are the plans for the startup now?* We asked the executives about their view on the case startups (*How are the startups supported by Iterate*, *what is important for the startups to succeed? What are the obstacles that you encounter?*).

**Table 2.** Data sources

| Startup | Source ID | Source(s) description | Participant ID | Source type | Length of the interview (min) |
|---|---|---|---|---|---|
| 1 | I1 | Startup founder | S1P1 | Semi-structured interview | 60 |
| 2 | I2 | Startup founder | S2P1 | Semi-structured group interview | 85 |
| N/A | | Senior Executive | E1 | | |
| 3 | I3 | Senior Designer in a startup team | S3P1 | Semi-structured interview | 75 |
| | I4 | External program manager | S3P2 | Informal interview | 40 |
| | I5 | External program manager | S3P3 | Informal interview | 30 |
| 4 | I6 | Startup founder | S4P1 | Semi-structured interview | 55 |
| N/A | I7 | Senior Executive | E1 | Semi-structured group interview with the executives | 85 |
| | | Senior Executive | E2 | | |
| N/A | M1 | Kick-off of the research project | E1 | Meeting notes | 120 |
| N/A | M2 | Presentation of the preliminary results from the data analysis | E1, E2 | Meeting notes | 30 |
| N/A | M3 | Presentation of the first paper draft | E1, E2 | Transcript of the video recording | 67 |
| Documents: Status emails, pitching slides, and webpages related to the startups | | | | | |

In this paper, we present results from the data analysis that was based on the procedures suggested by Hoda [27] in her Socio-Technical Grounded Theory (STGT) for Software Engineering. These procedures are recommended for the studies applying the Grounded Theory approach but are also described as suitable for other types of studies, e.g. case studies as the current one [27]. The purpose of STGT is to set a methodological ground for studying both social and technological aspects of software engineering. This was crucial for our study because understanding internal software startups implies both understanding the technology under development and the social aspects (such as interaction within the startup teams or with the parent company). *Literature review*, *Open coding*, *constant comparison*, *basic memoing*, and *axial coding* with the use of a pre-defined theoretical template, are the procedures that we applied, as we proceed to describe. *Literature review* should be performed both early in the study (*lean literature review*) and periodically as the results mature (*targeted literature review)* [27]. We performed a *lean literature review* prior to the data analysis to identify the research gaps. The *targeted review* was conducted during the preparation of this manuscript to position the results within the existing research on internal startups and intrapreneurship. *Open coding* is a process of representing textual raw data into a condensed format and where all the data is covered [27]. To perform open coding, we inserted all the interview transcripts into the analytical tool NVivo 12 and coded all instances in relation to our research question (see Table 3). The codes represented a phrase, and each phrase summarized an abstract or several sentences. Following the *constant comparison* technique, the codes were continuously compared within and across the data sources, to identify meaningful patterns [27]. *Concepts* are patterns of codes at a lower level of abstraction, whereas *sub-categories* and *categories* – at higher levels of abstraction [28]. To reflect on the emerging data structure, the first author applied *basic memoing*, which is a technique to document the researcher´s thoughts on the emerging concepts and categories [27]. Small texts (memos) were written down and updated throughout the analysis process to describe the concepts and why they include the particular codes. The data collection was carried out in parallel with the data analysis, which allowed for iterative data analysis when the emerging concepts and categories informed the latest data collection.

A pre-defined theoretical template from Masood et al. [28] was applied as an analytical tool to map the relationships between the emerging concepts and categories. Such templates are recommended by STGT to guide the theory development when the examined phenomena are relatively narrow (as in our case "How does a venture builder make internal startups work?") [27]. The template is one way to systematically relate categories and sub-categories, which is the essence of *axial coding* [28]. The application of the template allowed us to map the emerging concepts and categories to the pre-defined fields, e.g. strategies, intervening conditions, facilitating conditions, consequences, etc. This paper reports mainly from the category identified as *strategies* and partly from the category *consequences* that together reflect how the case company makes its internal startups work. In doing so we followed the recommendations by Hoda [27] to report parts of an emerging grounded theory (e.g. individual categories) throughout the process of the theory development. The purpose of this is to acquire feedback from practitioners and the research community to guide further theory development.

**Table 3.** Examples of applying the STGT analytical procedures

| Interview transcript | Open coding | Concepts | Category |
|---|---|---|---|
| *"We have what we call Iterate Time. We can use a certain number of hours per month to learn something new or to do something one is really excited about. So we decided to use this time to work until the next sprint"* - S1P1 | Using Iterate Time to finance work in the startup teams | Iterate Time, Startup teams | Strategies; **sub-category**: financial |
| "*We are a highly autonomous team [...], we take responsibility for the tasks that we choose and never must ask each other "Now you should remember to do* | Culture in the startup team has everything to do | Shared culture, Autonomy | Consequenc es; **sub-category:** |
| *this*". *This culture has everything to do with the culture at Iterate"* – S1P1 | with the culture in the parent company | , Startup teams | startup teams |
| "*The fact that the employees own the company as a whole, makes us all wanting to help each other*" – E2 | All employees have a share in the parent company to motivate everyone to help each other | Employees and investors and shareholders | Strategies; sub-category: financial |

# 4 Results

We will now present the seven identified strategies that appeared crucial for the internal software startups to work (identified consequences) at Iterate (Table 4). The strategies were grouped into four sub-categories (cultural, financial, personnel, and venture arrangement).

**Table 4.** Overview of the strategies

| Strategy type | Description | Identified consequences |
|---|---|---|
| Cultural | **S1**: Exposing employees to each other | Motivation to formulate new ideas, source of feedback for the initial startup ideas, networking among the employees, a lower threshold to ask for help<br>Employees become capable of starting up their own businesses |
| | **S2**: Creating implicit norms of taking initiative | Autonomy and responsibility in the startup team |
| Financial | **S3**: Incremental funding | Employees have time to explore their ideas, the possibility to recruit others to the startup teams, time squeeze |
| | **S4**: Shared incentives | Willingness to collaborate, committed startup teams over time |
| Personnel | **S5**: Highly selective recruitment | Employees´ readiness for entrepreneurship |
| | **S6**: Executives as coaches | Startups receive support, executives make informed decisions on further investment |
| Venture arrangement | **S7**: Postponing distribution of shares | Extended experimentation period, the equity is shared among the most committed team members |

## 4.1 Cultural strategies

The two identified cultural strategies reflect how Iterate promotes a climate meant to create favorable conditions for internal startups.

**S1: Exposing employees to each other.** The idea behind Iterate is to create new ventures based on the ideas of the employees. Therefore, it is essential to promote an environment where ideas emerge and grow. Iterate has established arenas where employees pitch ideas, give and receive feedback on ideas and collaborate on developing new ideas. The company does it by promoting social events that expose employees to each other and stimulate networking. The executives emphasized that they strive to lighten up the company through several common events that happen on a regular basis (see Table 5). One of them said: "*We make sure that these events are the beating heart of the company. We offer a diversity of meeting places where people can get together, learn from each other, and get inspired […] We do not manage the content in this, but we make sure that this happens"* - E2, *executive*. The employees described these arenas as engaging, motivating, and a useful source of feedback on their ideas. It appeared that the events led the employees to be well-acquainted with each other, which lowered the threshold for asking for help, as one startup founder described: "*Due to all the social stuff that we have at Iterate, we have met everyone that works at Iterate, so we know whom to ask if we wonder about something, and they help because they know us*" - S4P1, *Startup founder.*

**Table 5.** Overview of the main events

| Event | Description | Purpose | Frequency |
|---|---|---|---|
| Ship-It Day | Hackathon where people have 24 hours to work with totally new ideas or solve a complicated problem. The result should be a finalized prototype or a solution | Take a break from the routine, develop a somewhat finalized product | Annually |
| Breakfast meetings | An informal 30-minutes presentation around a breakfast table (before the pandemic) or digitally (during the pandemic). The presenter is randomly chosen at the beginning of the meeting | People get insight into what others are working with, training in spontaneous presentation, an arena for new startup teams to emerge | Flexible |
| Pitch Night | Everyone is encouraged to pitch all kinds of ideas, no matter how good or bad. | Lower the threshold for how good a new idea can be, have fun, share inspiration and ideas, an arena for new startup teams to emerge | Monthly |
| While-We-Wait | A mini-conference where people with product management experience share it with the others | Increase people´s competence in product management | Weekly |

**S2: Creating implicit norms of taking initiative.** Being initially an IT consultancy company, the transition to becoming a venture builder required a cultural shift for people from solving technical problems to being entrepreneurs. To promote this shift, Iterate trains people at making independent decisions and taking initiative, which people also call "actionocracy". One of the executives explained that if employees ask to purchase new furniture or other things for the office, he encourages them to do it themselves. He said: "*If you cannot buy new furniture for the office, then you cannot build a new company, either*" – E1, *executive*. The norms have an implicit nature, because they stem from people´s behaviors and unspoken assumptions, not from written guidelines. Different participants referred to these norms as "invisible principles" (S2P1), "Iterate DNA" (S1P1), and "Iterate culture" that all point out to this implicit nature.

The norms are promoted by numerous means. Even before the actual employment, the norms are shared through student meetings, summer internships, and even through informal networks of the current employees. The norms are reinforced through the social events (see S1) and supported by the close interaction with the executives (S7) who themselves serve as good examples of taking initiative. Arrangements such as Ship-It Day (Table 5) and Iterate Time (S3) also nudge the employees towards initiating and completing the tasks that they themselves find meaningful. Ultimately, the norms support employees in becoming startup founders. In addition, they promote a high level of both autonomy and responsibility in the startup teams, as indicated by this quote: "*We are a highly autonomous team […], we take responsibility for the tasks that we choose and never must ask each other "Now you should remember to do this*". *This culture has everything to do with the culture at Iterate"* – S1P1, *startup founder*.

### 4.2 Financial strategies

The two financial strategies create conditions for financing the startups in the initial phase and attracting investors for the future.

**S3: Incremental funding.** The IT consultancy is a crucial source of revenue for Iterate, but it is also important to allow people to explore their own ideas. To balance these two, Iterate has established flexible mechanisms for funding to increase incrementally as ideas mature. The initial funding is enabled through about 10-20% work time that the employees are free to use for everything they find interesting, important, and meaningful. At Iterate it is known as "Iterate Time" and can in principle be used for absolutely anything. Many startup founders (such as S1P1, S2P1) used Iterate Time to work on ideation and initial user testing of their startup ideas. When 10-20% work time becomes insufficient for running a startup, the founders may request additional funding in form of hours paid for a month of work for the startup team or renting other resources from Iterate (e.g. UX-designer, additional developers). An executive, who is involved in decisions on the funding, commented: "*These are very easy and minor decisions for me because I don´t fund the whole course, but just the time that costs much less. And we do this all the way*", - E2, executive.

Incremental funding allows the startup teams to finance their members´ work up until (and if) they manage to acquire external funding at later stages. Iterate Time increased the number of startup ideas because many employees had resources to initiate their startups. It also created opportunities for other employees to contribute to the emerging ideas as they could "donate" their Iterate Time by working in other startup teams. On the other hand, some founders struggled in the initial phase because Iterate Time was not always sufficient to perform all the work that the startup founders would like to do, which made some of them squeezed between the startup and the consultant tasks, as shown in this excerpt: "*I would like to have an extra day to do all the things that I prioritize [in the startup] and rather go down 20% in salary to make up for it […] but I would also like to continue in my daily consultant work*" - S2P1, *startup founder*.

**S4: Shared incentives.** The business model of creating new ventures is directly dependent on the employees´ willingness to become venture builders. To motivate them, Iterate promotes shared incentives for employees to avoid the competition created by individual bonuses. Being an employee-owned organization Iterate encourages its employees to also be its shareholders so that if Iterate´s market value increases, everyone will benefit. According to the executives, this creates incentives for mutual collaboration, as indicated in this quote: *"The fact that the employees own the company as a whole, makes us all wanting to help each other. No individual bonuses at any level, because this creates conflicts"* – E2, *executive*. Apart from being shareholders of the parent company, the employees can also invest money in individual startups. This is mutually beneficial, because the startups acquire additional funding, whereas the internal investors acquire potential financial benefits. When people invest in the startups of others, they may also be motivated to help "their" startup team succeed.

Further, the startup founders own the majority of shares in their own startups, whereas Iterate holds the rest of the shares. This secures the commitment and responsibility of the startup team over time. One of the executives expressed: "*Owning shares gives the feeling of responsibility for caring about the startup over time, and this should belong to the team. Therefore, Iterate owns only 30%, and the team owns 70%*» - F2, *executive.* Since the startup teams own most of the startup many team members are willing to put extra work into the startups. Many startup founders were committed and worked in their spare time, as illustrated in this quote: "*I do not see the startup job as a job. I mean, there is some work in your spare time because it is fun to be able to build your own future*" - S2P1, *startup founder*.

## 4.3 Strategies related to personnel

The following two strategies reflect how the company recruits people and how people are supported in being entrepreneurs.

**S5: Highly selective recruitment.** People employed at Iterate should collectively possess skills that make them capable of developing new products internally. However, having technical skills, such as programming or designing, is not sufficient. Therefore, Iterate invests a lot in the recruitment process to find people who are both technically strong, as well as open to becoming entrepreneurs. A lot of today´s recruitment happens among the students through a Summer internship. The highly competitive process (only 9 candidates of 200 receive the internship offer) takes up to 2 years until the candidate receives a permanent job offer. According to the executives, it is extremely important to find people that have an open mindset (e.g., among newly-graduated) and are robust in the face of uncertainty and effort of the entrepreneurship. One of them said: "*The* *type of people who aspire to be entrepreneurs but also understand how much effort it takes. These are the perfect candidates for us*" - E2, *executive*.

**S6: Executives as coaches and co-founders.** The majority of employees lack experience in establishing a new business/startup. Therefore, the executives offer coaching to those who work with new ideas. The coaching addresses issues as finding matching venture capitalists (S2P1, E1), distribution of shares between the members of the startup team (E2), and general product management (E2). A lot of coaching and feedback is informal since the executives often are also board members in the emerging startups, which allows them to participate in the decision-making process together with the startup teams. The startup founders themselves decide whether and when to ask for the coaching. One of the executives pointed out: "*If people feel that they need to validate things with us before they start, it will go too slowly. But when things are starting to become interesting, we are happy to be in the dialog*" – E1, *executive*.

The coaching is beneficial for both the startup founders who receive guidance and support and the executives who are thus continuously updated on the startups´ progress and challenges. Working closely with the startups helps the executives to calibrate the startup´s funding if necessary (see S3: Incremental funding). Further, the level of personal engagement in the startups is a good indicator of future success; one executive said: "*The more we coach them, the more we know them, whether they are passionate and whether they learn. And these are passion and learning we look for when we take bigger decisions on investment*" – E1, *executive.* The startup founders regarded the coaches as important, available, and motivating. One of them described: "*Our manager is very adept at showing how he can help. He's a busy man but always seems to have time for us […] He is also a great motivator*" – S4P1, *startup founder.*

### 4.4 Strategies related to venture arrangement

In this category, we describe one strategy that explains how Iterate deals with internal startups during the transition to a spin-off phase.

**S7: Postponing distribution of shares.** Since the team members work on the startup as long as they find it meaningful, they can leave at any time, whereas other members can come in. Besides, different team members contribute to the startups differently. To account for this, Iterate advises the startup teams to postpone the official registration of the new companies until the team has been stable for about 1-3 years. One executive noted: "*Those who want to start up together must date each other so long that they feel like getting married. […] Also, it is very important to continuously discuss how the shares should be distributed, it is very important*" – E2, *executive*. Postponing the distribution of shares enables the startup teams to extend the prototyping phase while at the same time having flexibility in the team composition. A startup founder explained: "*When we missed people from the team, we still had not distributed the shares. Since we had not registered the company by then, we did not have to do* *emissions, to buy and sell again. We had only oral agreements until a certain point, it was an advantage for us*“ - S1P1, *startup founder.*

## 5 Discussion and practical implications

We will now discuss our findings to answer the study´s research question ´*What makes internal startups work´?* The overall contribution of our study is an insight into intrapreneurship/internal venture [19] in the context of internal software startups in a venture building company.

Earlier research has demonstrated that intrapreneurship culture predicts the market performance of IT-enabled firms [29]. Our findings also show that intrapreneurship culture is a key for establishing internal software startups. We also found which exactly cultural strategies were beneficial for the startups in a venture builder environment. For example, by *exposing employees to each other* (S1), the case company strengthened networks among the employees, eventually enabling mutual collaboration, trust, and a shared spirit of entrepreneurship. Our findings indicate that when employees know each other well, they are likely to reach out for help thus acquiring access to the in-house expertise and resources. Moreover, informal networks facilitate the self-organization of the employees in the startup teams. Earlier research has indicated that access to the existing networks of experts internally in the company has a positive effect on internal startups [7]. Networking is also crucial for stand-alone software startups because it gives the founders access to people with different expertise and resources [18]. Therefore, establishing arenas for sharing ideas in an informal way (such as the Pitch Night) can be recommended to enable internal software startups.

From before, we know that internal venture and intrapreneurship activities are often financially constrained [19], which is why it is necessary to provide internal startups with sufficient funding. As appears from our results, financial strategies such as *incremental funding* (S3), is a good way to solve the problem of funding while at the same time keeping the startups´ autonomy. The startup founders that used Iterate Time could dedicate some hours to work with the idea without any control from the executives. Moreover, the founders could convince others to join their team by “donating” their Iterate Time, which gave autonomy in designing the team. However, merely providing employees with extra hours to experiment does not have to result in internal software startups. Therefore, the startup teams at Iterate received *incremental funding* (S3), which conveyed their responsibility over time and created continuity and commitment. Strategies similar to Iterate Time are used under different names also in other companies (e.g. “15% Time” in Atlassian [4] and M3 [5] ) and have two main benefits: 1) contribute to capturing, developing, and testing novel ideas 2) reduce the risks for the parent companies to invest in ideas that may not work. Our findings suggest that x% time strategies should be combined with incremental funding to extract the most potential out of internal software startups.

Further, earlier research suggests that intrapreneurs can be less assertive than a typical top manager, which might create problems as the products grow [19]. Our data also indicated that startups founders might need support from someone more experienced in business, particularly, the executives who functioned as coaches (S6). Apart from being business mentoring, coaching also functioned as a motivational incentive and continuous informal monitoring and feedback. This strategy also helped the executives make case-by-case decisions on how the startups should be funded, in collaboration with the startup teams. Earlier research has indicated that top management support, such as that of the Iterate executives/coaches, speeds up the developments process of the internal startups [7]. We, therefore, recommend the coaching of the startup teams to be performed by experienced entrepreneurs with authority at the top level of the company.

Finally, increased personal commitment and emotional investment are common with intrapreneurship and internal ventures [19]. In line with this, we discovered that harnessing peoples´ own motivation to develop their ideas, was crucial to make the internal software startups work. Our results demonstrate that several strategies at the case company aimed at increasing the employees´ personal stake in the outcome of the startup. Through *highly selective recruitment* (S5), Iterate was selecting people that were motivated to develop their own business. By creating *shared incentives* (S4) of their startups and of Iterate, the company built up peoples´ financial interest in the outcome of the startups while at the same time fostering collaboration. Since the employees through *incremental funding* (S3) had room to work with what they were really dedicated to, they were gradually becoming emotionally engaged in their own projects to the extent that they were willing to work without being paid. *Postponing distribution of shares* (S7) ensured that the equity was distributed among those startup team members that had shown the most commitment over time. Earlier findings indicate that personal stake in the outcome strengthens the motivation of the internal software startup teams [7]. It is further proposed that equity offers both financial and emotional motivation to the startup team members [18]. Our results support these propositions also for internal software startups, indicating that personal stake in the outcome (e.g. making your own business succeed) is the main motivation of the startup founders and team members. Therefore, we recommend parent companies identify the people motivated for entrepreneurship and support their motivation by various means (e.g. financial incentives, work hours, recognition of effort) while at the same time promoting collaboration. The desire to make one´s own idea work should be recognized as the main driver behind internal software startups.

## 6 Limitations, conclusions, and further work

This study provides several insights into how the internal software startups are enabled at a venture building company Iterate, which is a particular type of company.

The findings should therefore not be generalized to other contexts without precautions. What works in a relatively small venture-building company like Iterate does not necessarily work in a large-scale environment.

Taking the limitations into account, we conclude that the strategies that companies might apply to make internal software startups work, contribute to 1) more autonomous and collaborative organizational culture, 2) continuous financial resources for the startups 3) in-house competence in product management, and 4) flexible ownership of the emerging startups. Based on the findings, we recommend that organizations who wish to succeed with internal software startups 1) establish arenas where employees can share ideas and build informal networks, 2) provide incremental funding to experiment and establish self-managing startup teams, 3) arrange the startup teams to be coached by experienced entrepreneurs with authority in the company, and 4) identify people motivated for entrepreneurship and harness their motivation.

Further research should address the question of which innovation strategies are/should be used in contexts other than those of venture builders (e.g. in large-scale companies). We also aim to develop a more refined grounded theory that explains how the strategies for promoting internal software startups vary depending on their context and level of maturity. Finally, there is a need for a more theoretically grounded approach to understanding internal software startups.